# A $Fe^{3+}$-hydroxide Ligation in the Superoxide Reductase from *Desulfoarculus baarsii* is Associated with pH Dependent Spectral Changes.

*Christelle Mathé,*[†, §] *Vincent Nivière,*[*, §] *and Tony A. Mattioli* [*, †]

† Laboratoire de Biophysique du Stress Oxydant, SBE/DBJC and CNRS URA 2096, CEA/Saclay, 91191 Gif-sur-Yvette cedex, France, and § Laboratoire de Chimie et Biochimie des Centres Redox Biologiques, DRDC-CEA/CNRS/Université J. Fourier, UMR 5047, 38054 Grenoble Cedex 9, France.

AUTHOR EMAIL ADDRESS: tony.mattioli@cea.fr; vniviere@cea.fr

TITLE RUNNING HEAD: $Fe^{3+}$-OH Species in Superoxide Reductase Alkaline Transition

**ABSTRACT**

Superoxide reductase (SOR) catalyzes the reduction of $O_2^{\bullet-}$ to $H_2O_2$. Its active site consists of a non-heme $Fe^{2+}$ center in an unusual square-pyramidal [$His_4$ Cys] coordination. Like many SORs, the electronic absorption band corresponding to the oxidized active site of the SOR from *Desulfoarculus baarsii* exhibits a pH-dependent "alkaline transition" changing from *ca*. 644 to 560 nm as pH increases and with an apparent $pK_a$ of 9.0. Variants in which the conserved amino acids glutamate 47 and lysine 48 were replaced by the neutral residues alanine (E47A) and isoleucine (K48I), respectively, exhibited the same alkaline transition but at lower apparent $pK_a$ values of 6.7 and 7.6, respectively. Previous work [Nivière, V.; Asso, M.; Weill, C. O.; Lombard, M.; Guigliarelli, B.; Favaudon, V.; Houée-Levin, C. *Biochemistry* **2004**, *43*, 808-818] has shown that this alkaline transition is associated with the protonation/deprotonation of an unidentified base, $B^-$, which is neither E47 or K48. In this work, we show by resonance Raman spectroscopy that at basic pH a high-spin $Fe^{3+}$-OH species is formed at the active site. The presence of the $HO^-$ ligand was directly associated with an absorption band maximum at 560 nm, whereas upon protonation the band shifts to 644 nm. With respect to our previous work, $B^-$ can be identified with this high-spin $Fe^{3+}$-OH species, which upon protonation results in a water molecule at the active site. Implications for the SOR catalytic cycle are proposed.

**Introduction**

Superoxide reductase (SOR) catalyzes the one-electron reduction of $O_2^{\bullet-}$ to $H_2O_2$ providing an antioxidant defense in some anaerobic or microaerophilic bacteria:[1-4]

$$O_2^{\bullet-} + 1e^- + 2H^+ \rightarrow H_2O_2$$

Historically, the only enzymes which were known to eliminate superoxide were superoxide distmutases (SODs) which catalyze the disproportionation of superoxide.[5] The recently discovered SOR activity involving the reduction of superoxide has changed our view of how toxic superoxide is eliminated in cells and how this activity protects the cell against oxidative stress and lethal levels of superoxide. Consequently, this new enzymatic activity has stimulated interest in the synthesis of chemical model systems in attempts to mimic the SOR reaction.[6-9]

SORs are a novel class of non-heme iron proteins that can be classified into one-iron proteins, which possess only the active site, and two-iron proteins which possess an additional rubredoxin-like $Fe^{3+}$-$(SCys)_4$, center for which the function and role are not known.[10-12] In the reduced state, the SOR active site consists of a non-heme $Fe^{2+}$ center in an unusual [His$_4$ Cys] square pyramidal pentacoordination. The sixth coordination site of the $Fe^{2+}$ center is vacant and suggests the most obvious site for $O_2^{\bullet-}$ binding.[13-15] The SOR active site $Fe^{2+}$ center reacts specifically at nearly diffusion-controlled rates with $O_2^{\bullet-}$ according to an inner sphere mechanism.[16-19] Although one or two reaction intermediates have been proposed depending on the enzyme studied, it is now generally accepted that these intermediate species are $Fe^{3+}$-peroxo species.[16,19-22] For the *Desulfoarculus baarsii* enzyme, the reaction mechanism was proposed to involve formation of two intermediates.[16,19] The first one, a proposed $Fe^{3+}$-peroxo species, results from the bimolecular reaction of $O_2^{\bullet-}$ with $Fe^{2+}$. It then undergoes a diffusion limited protonation process to form a second reaction intermediate, a $Fe^{3+}$-hydroperoxo species.[19] A second protonation process at the level of the $Fe^{3+}$-hydroperoxo species would allow the release of the reaction product $H_2O_2$.[19] Although the donor of this second protonation event is not known yet, it was proposed to be associated with the presence of a protonated base BH at the SOR active site.[19] Finally, the resulting $Fe^{3+}$ atom of the active site becomes hexacoordinated with a conserved glutamate residue (E47 in *D. baarsii*).[14,23,24]

A recent investigation of pH effects on the *D. baarsii* SOR active site revealed that the redox midpoint potential and absorption spectrum of the oxidized active site both exhibit similar pH transitions.[19] The active site absorption band at 644 nm, which arises from a CysS → high spin $Fe^{3+}$ charge transfer (CT) transition,[24] shifted to 560 nm as pH increases, with an observed $pK_a$ of 9.0.[19] No such pH dependent absorption changes were seen for the other rubredoxin-like $Fe^{3+}(Cys)_4$ center of the *D. baarsii* SOR.[19] EPR spectroscopy showed that, like the 644 nm-absorbing species, the 560 nm species also arises from a high spin $Fe^{3+}$, but with a modified rhombic signal at g = 4.3,[19] suggesting a possible change in iron coordination. The data were fully consistent with the presence of a base (B⁻) in the active site, which modulated its redox and spectral properties. When the base is in its deprotonated form (B⁻), reduction of the iron active site involves both one proton and one electron, and hence the redox potential is pH dependent.[19] When the base is in its protonated form (BH), the reduction of the active site involves only one electron and the redox potential is pH independent.[19] The presence of a base, responsible for a similar spectroscopic alkaline transition, has also been described for the SORs from *P. furiosus* and *D. gigas*.[25,26] However, in no case has this base been identified.[19,25,26] For the case of the *P. furiosus* SOR, resonance Raman (RR) experiments at different pH values were inconclusive concerning the origin of the alkaline transition and provided no evidence for the formation of a ligand trans to the cysteine residue at high pH.[25] Finally, the base was shown to be influenced by the two closest charged residues to the active site, E47 and K48, as the mutation of these two residues (E47A and K48I) modified the $pK_a$ value of the pH transition described above to 6.7 and 7.6, respectively.[19] These two residues are conserved among SORs and are proposed to play essential roles in SOR activity. This has been clearly demonstrated for the lysine residue, which plays an important role for electrostatic attraction of superoxide.[16,19] Based on low temperature trapping experiments where $H_2O_2$ was reacted with SOR, the glutamate residue was proposed to facilitate $H_2O_2$ release.[20] However, during the reaction with $O_2^{\cdot-}$ such a role has not been demonstrated yet.

Protonation events play crucial roles for the SOR reaction since two protons are required for the formation of $H_2O_2$ from $O_2^{\cdot-}$. In this work, we have investigated the pH-dependent structural and coordination changes of the active site of the *D. baarsii* SOR using resonance Raman spectroscopy. We demonstrate here that the species associated with base B[-] and which is directly responsible for the pH-dependent spectral and redox changes mentioned above[19] is a previously unidentified high-spin $Fe^{3+}$-OH species. Taken together with recent previous work,[19] the results presented here indicate that B[-] is HO[-] and suggest that the donor BH of the second protonation event is a water molecule at the active site. The functional implications for the SOR mechanism are discussed.

**Experimental Procedures**

*Biochemical and chemical reagents*. $K_3Fe(CN)_6$, ammonium persulfate were from Sigma. $K_2IrCl_6$ was from Strem Chemical Inc. $^2H_2O$ was purchased from Aldrich and $H_2^{18}O$ was purchased from ICON Stable Isotopes.

*Protein purification.* Purification of the wild-type, E47A and K48I proteins from *D. baarsii* were performed as described elsewhere.[2,16] Purified protein samples, in 10 mM Tris-HCl pH 7.6, were concentrated with Microcon 10 microconcentrators (10 kDa cut-off membranes, Amicon). Unless otherwise stated, proteins were oxidized using 3 equivalents of $K_2IrCl_6$. In some cases, 1.5 equivalents of $K_3Fe(CN)_6$ or 1 equivalent of ammonium persulfate were used. In general, the excess oxidant was removed by washing using Microcon 10 microconcentrators.

SOR protein samples were poised at different pH values using the following buffers: pH 5, 150 mM acetate; pH 7.0, 150 mM phosphate buffer; pH 7.5–8.5, 100 mM Tris-HCl; pH 10, 100 mM glycine/NaOH. Samples were verified to be within 0.1 pH value before and after freezing.

*Spectroscopy.* Optical absorption measurements were made with a Varian Cary 1 Bio spectrophotometer, with protein samples in 1 cm path length cuvettes. Low temperature (4.2 K) X-band EPR spectra were recorded with a Bruker EMX 081 spectrometer equipped with an Oxford Instruments continuous flow cold He gas cryostat.

In general for the resonance Raman samples, 1 µl of concentrated protein (1-5 mM) in ca. 100 mM of buffer was deposited on to a glass slide sample holder and then transferred into a cold helium gas circulating optical cryostat (STVP-100, Janis Research) held at 15 K. The pH of trial samples (200 µL, 100 mM buffer) were re-measured after freezing/thawing and were observed to not change within ± 0.1 pH unit.

Resonance Raman spectra were recorded using a modified single-stage spectrometer (Jobin-Yvon) equipped with a liquid nitrogen-cooled back-thinned CCD detector (2000 X 800 pixels), and excitation at 647.1 nm (30 mW) was provided by a Spectra Physics Series 2000 $Kr^+$ laser. Stray light was rejected with a holographic notch filter (Kaiser Optical). Spectra were calibrated using the exciting laser line, along with the $SO_4^{2-}$ (983 $cm^{-1}$) and ice (230 $cm^{-1}$) Raman bands from a frozen aqueous sodium sulphate solution. Spectral resolution was < 3 $cm^{-1}$ (entrance slits at 100 µm). Frequency accuracy was ± 1 $cm^{-1}$ and frequency repeatability was 1 pixel (resolution < 2 pixels) for the spectrograph. Reported spectra were the result of the averaging of 40 single spectra recorded with 30 seconds of CCD exposure time; no spectral smoothing was performed. The reproducibility of the reported Raman frequencies was determined to be ± 1 $cm^{-1}$. Verification of small frequency differences of 2 $cm^{-1}$ of homologous bands originating from two different samples was established by recording, during the same day, the spectra of the two different samples deposited on the same sample holder under the same spectroscopic/geometric conditions. Baseline corrections were performed using GRAMS 32 software (Galactic Industries). In all reported spectra, the Raman contributions from ice have been subtracted using the GRAMS 32 software.

**Results and Discussion**

The SOR from *Desulfoarculus baarsii* possesses two non-heme iron binding sites.[2,15] For the "as-isolated" protein, the active site iron center is in the reduced $Fe^{2+}$ state and exhibits no electronic absorption bands in the 500-800 nm region. The other iron site, the rubredoxin-like non-heme site not participating in SOR chemistry, is in the oxidized $Fe^{3+}$ state and exhibits an electronic absorption band at ca. 500 nm arising from $S \rightarrow Fe^{3+}$ charge transfer bands from that iron center.[2] After treating the "as-isolated" SOR with a non-complexing oxidant such as $K_2IrCl_6$, the active site iron becomes oxidized and exhibits an absorption band at ca. 650 nm at pH 7.5.[19,24] This absorption band is due to a $S \rightarrow Fe^{3+}$ charge transfer band of the active site.[24] Therefore, the resonance Raman spectra of the fully oxidized *D. baarsii* SOR (i.e. both iron centers in the $Fe^{3+}$ redox state) excited using a 647.1 nm excitation, will exhibit predominant resonant contributions from the oxidized active site $Fe^{3+}$-S(Cys) moiety[20,25,27] as well as relatively weaker preresonance contributions from the $Fe^{3+}(SCys)_4$ rubredoxin-like site, as previously discussed.[20] **Figure 1** shows the RR spectrum of the oxidized *D. baarsii* WT SOR active site at pH 7.5, where the preresonance Raman contributions of the rubredoxin-like site have been subtracted.

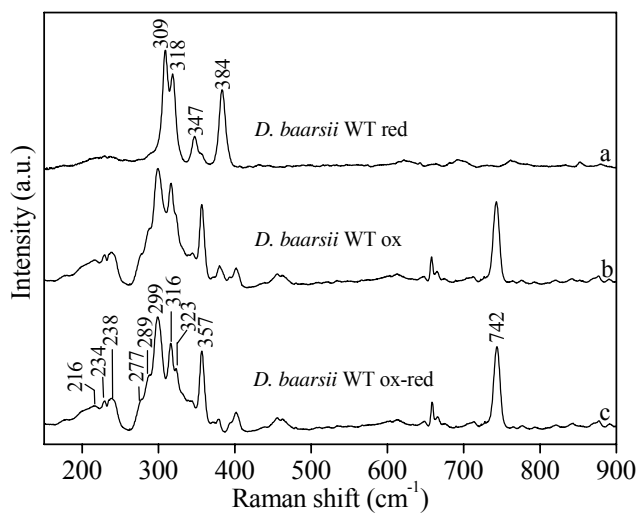

**Figure 1.** Subtraction of the rubredoxin-like iron center pre-resonance Raman contributions from the low temperature (15 K) resonance Raman spectra of WT SOR from *D. baarsii* (3 mM in 100 mM Tris-HCl pH 7.5). (a) untreated (active site reduced), (b) active site oxidized with 3 eq $K_2IrCl_6$, and (c) the calculated difference of Spectrum (a) subtracted from Spectrum (b) with the cancellation of the 384 cm$^{-1}$ band arising uniquely from the rubredoxin-like center. Spectra (a) and (b) are not shown at the same scale. The rubredoxin-like center remains in the same $Fe^{3+}$ state in (a) and (b). Spectrum (a) shows the pre-resonance Raman spectrum of the rubredoxin-like center with prominent bands at 309, 318, 347, and 384 cm$^{-1}$. Excitation wavelength was 647.1 nm, 50 mW of laser power.

The RR spectra of several SOR active sites have been reported and the various RR bands have been tentatively assigned based on global S- and N-labelling of the proteins.[25,27] The RR spectrum of the oxidized active site of the SOR from *D. baarsii* exhibits bands (**Figure 1**) similar to those described for *P. furiosus* and *D. vulgaris* Hildenborough.[25,27] In the low frequency region, the prominent bands at 299, 316, and 232 are due to coupled Fe-S stretching and bending modes, while the band at *ca.* 742 cm$^{-1}$ is predominantly due to the C-S stretching mode of the cysteine ligand (**Figure 1**).

**Figure 2A** shows the RR spectrum of the oxidized active site of the *D. baarsii* WT SOR at pH 8.5 and at pH 10 where the absorption band shifts from 644 nm to 560 nm, respectively (**Figure 2B**). At pH 10, the high spin Fe$^{3+}$ EPR spectrum of the active site exhibits modifications.[19] Comparison of these two RR spectra (**Figures 2A,a and 2A,b**) reveals a new band at 466 cm$^{-1}$ for the "560 nm-absorbing" species (pH 10) which is not seen for the "644 nm-absorbing" species (pH 8.5). The RR new band is also seen for the 560 nm-absorbing form of both the K48I (471 cm$^{-1}$, **Figure 2A,c**) and the E47A (467 cm$^{-1}$, **Figure 2A,d**) variants, albeit with slightly different frequencies. We stress that the new 466-471 cm$^{-1}$ band is not observed for the respective 644 nm-absorbing species of either variant (**Figure 3**).

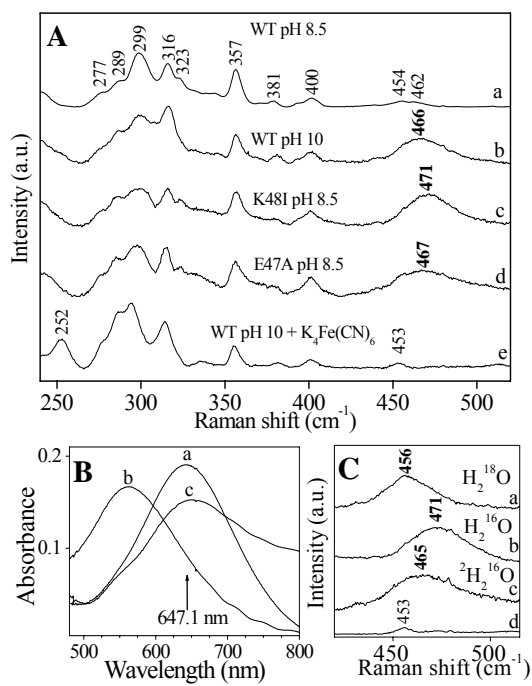

**Figure 2**. **A.** Low temperature (15K) resonance Raman spectra of the SOR active site for wild-type, K48I and E47A variants from *D. baarsii* (~4 mM in 150 mM buffer) oxidized with 3 eq of K$_2$IrCl$_6$. (a) Wild-type in Tris-HCl pH 8.5. (b) Wild-type in glycine/NaOH pH 10.0. (c) K48I variant in Tris-HCl pH

8.5. (d) E47A variant in Tris-HCl pH 8.5. (e) the same as (b) plus 5 eq of $K_4Fe^{2+}(CN)_6$. **B.** Room temperature absorption spectra of SOR active site wild-type from *D. baarsii* (100 μM) oxidized with 3 eq of $K_2IrCl_6$ at (a) pH 7.5 in 50 mM Tris-HCl, (b) pH 10.0 in 50 mM glycine/NaOH, (c) same as (b) plus 9 eq of $K_4Fe^{2+}(CN)_6$. **C.** Resonance Raman (15K) spectra of SOR active site K48I variant from *D. baarsii* at pH 8.5; same conditions as in Figure 1A. (a) oxidized with 3 eq $K_2IrCl_6$ in $H_2^{18}O$ buffer, (b) oxidized with 3 eq $K_2IrCl_6$ in $H_2O$ buffer, (c) oxidized with 3 eq $K_2IrCl_6$ in $^2H_2O$ buffer, (d) oxidized with $K_3Fe^{3+}(CN)_6$ in $H_2O$ buffer. In A, B and C, the contribution of the $Fe^{3+}$ rubredoxin-like center was subtracted from each spectrum. Same resonance Raman conditions as in Figure 1.

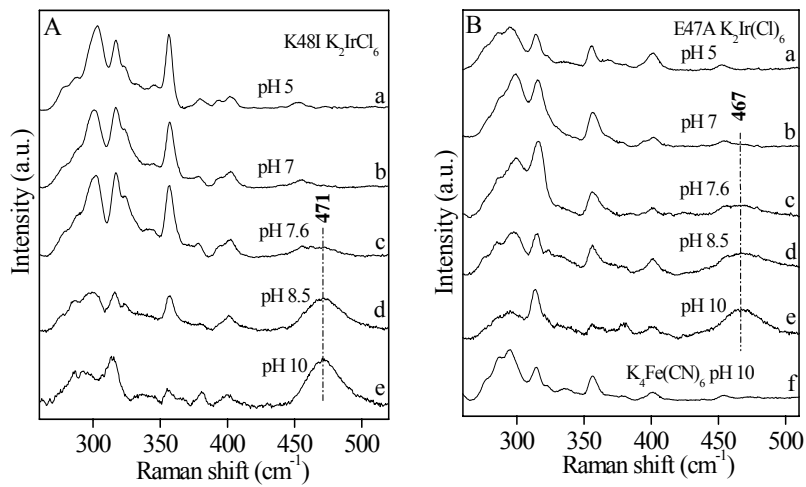

**Figure 3.** pH dependence of the low temperature (15 K) resonance Raman spectra of (A) K48I and (B) E47A variants from *D. baarsii* SOR (~4 mM) oxidized with 3 eq of $K_2IrCl_6$. Same Raman conditions as in Figure 1. The contributions of the $Fe^{3+}$ rubredoxin-like center was subtracted from each spectrum. (a) in 150 mM acetate buffer pH 5.0. (b) in 150 mM phosphate buffer pH 7.0. (c) in 150 mM Tris-HCl pH 7.6. (d) in 150 mM Tris-HCl pH 8.5. (e) in 150 mM glycine/NaOH pH 10. Spectrum (B,f) same as (B,e) plus 5 eq of $K_4Fe^{2+}(CN)_6$. Changes in relative intensities of the bands in the 250-350 cm$^{-1}$ range as pH increases reflect the changes in resonance Raman enhancement conditions, i.e. laser excitation remains at 647.1 nm but the active site absorption band changes from ca. 644 nm to 560 nm as pH increases.

While the new 466 cm$^{-1}$ RR band is distinctly observed for the WT at pH values greater than 8.5, the corresponding new band is seen to appear at lower pH values for the K48I and E47A variants (**Figure 2**). These observations are consistent with the reported apparent $pK_a$ values of the alkaline optical transitions of these SOR proteins ($pK_a$ = 9.0, 7.6, and 6.7 for WT, K48I, and E47A, respectively).[19]

The appearance of the 466-471 cm$^{-1}$ band for the *D. baarsii* WT, K48I and E47A SORs is completely pH-reversible for all three cases. **Figure 3** shows representative RR spectra of the SOR proteins at alkaline pH and after re-acidification. The new 466-471 nm band disappears when the solution is re-acidified. Furthermore, there are no changes in the frequencies of the other RR bands (**Figure 2A**; **Figure 4**), indicating no other observable secondary structural modifications or iron spin state changes for both the active site and the rubredoxin-like iron centers. These observations indicate that the appearance of the new 466-471 cm$^{-1}$ RR band is associated with a new ligand on the active site Fe$^{3+}$ center at high pH.

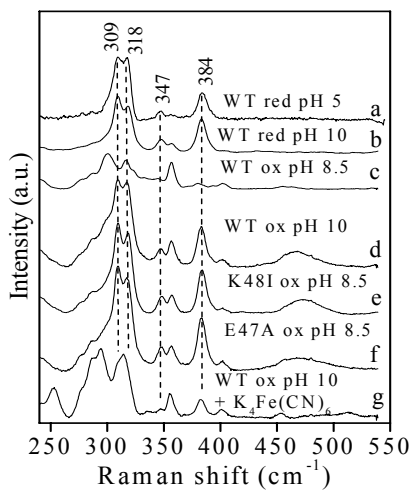

**Figure 4.** Low temperature (15 K) resonance Raman spectra of SOR from *D. baarsii*. Same Raman conditions as in Figure 2A. Trace (a) untreated WT *D. baarsii* SOR (1 mM) at pH 5 with active site reduced, trace (b) untreated WT *D. baarsii* SOR (4 mM) at pH 10 with active site reduced. The difference in signal-to-noise between spectra (a) and (b) is due to the lower concentration of protein in (a) and fewer spectral scans. Traces (c-g) same as Figure 2A (a-e) except that the pre-resonance contributions of the rubredoxin-like center were not subtracted here. The RR band frequencies of the rubredoxin-like center (309, 318, 347 and 384 cm$^{-1}$) do not change with pH and are the same for WT, E47A and K48I SORs. The 466-471 cm$^{-1}$ band assigned to the high-spin Fe$^{3+}$-OH stretching mode is clearly present in the absolute Raman spectra as it was in the difference Raman spectra of Figure 2A.

Oxidation of the SOR active site with K$_3$Fe$^{3+}$(CN)$_6$ is known to form a strong [$_{SOR}$Fe$^{3+}$-(NC)-Fe$^{2+}$(CN)$_5$] complex [15,28] exhibiting characteristic electronic absorption bands at ca. 650 nm and at ca. 1000 nm.[28] We observe that the 466-471 cm$^{-1}$ band does not appear when SORs wild-type, E47A (not shown) and K48I are oxidized with K$_3$Fe$^{3+}$(CN)$_6$ at high pH values (Figure 2C,d). In addition, for oxidized SOR solutions poised at pH values sufficiently alkaline to observe the new 466-471 cm$^{-1}$ band, the addition of K$_4$Fe$^{2+}$(CN)$_6$ completely abolishes the new band (**Figure 2A,e**). The absorption band observed at 560 nm at high pH also reverts back to a 644 nm band along with the appearance of the 1000 nm absorption band upon addition of $K_4Fe^{2+}(CN)_6$ (**Figure 2B**). This is totally consistent with the fact that the 466-471 cm$^{-1}$ RR band is associated with the 560 nm absorption band. When the SOR proteins are oxidized with $K_3Fe^{3+}(CN)_6$ at high pH values which would otherwise correspond to the appearance a 560 nm absorption band and the 466-471 cm$^{-1}$ RR band, the absorption band remains at 644 nm and the new 466-471 cm$^{-1}$ RR band is not observed. These experiments demonstrate that the new ligand can be easily displaced by the $Fe^{2+}(CN)_6^{4-}$ ligand.

## $^{18}O$ isotopic labelling

To identify the new ligand we have performed $H_2^{18}O$ and $^2H_2O$ experiments. Replacing the $H_2^{16}O$ buffer with $H_2^{18}O$ results in the observation of the new band at 456 cm$^{-1}$ instead of 471 cm$^{-1}$ as seen for the K48I SOR variant. No other RR band frequencies are affected (**Figure 5**).

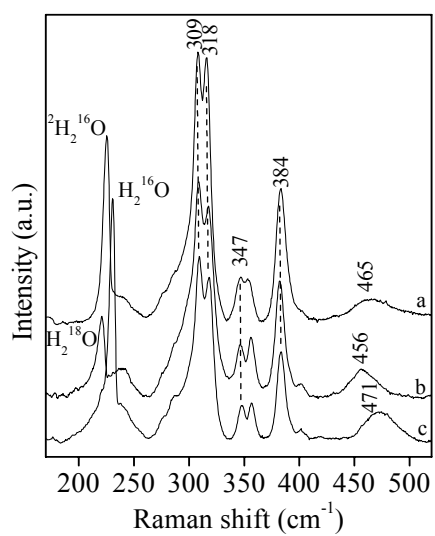

**Figure 5.** Low temperature (15 K) resonance Raman spectra of *D. baarsii* SOR K48I variant. Same Raman conditions as in Figure 2C. Spectra (a, b, c) same as in Figure 2C (c, a, b, respectively) except that the 200-500 cm$^{-1}$ region is now displayed. Absolute Raman data are shown where the pre-resonance Raman contributions of the rubredoxin-like center are seen at 309, 318, 347, and 384 cm$^{-1}$, indicating that these frequencies are unaffected by $H_2^{18}O$ and $^2H_2O$ exchange. Only the frequency of the $Fe^{3+}$-OH RR band (471 cm$^{-1}$) is affected by $H_2^{18}O$ and $^2H_2O$ exchange, as illustrated by the K48I variant.

The isotopic shift of -15 cm$^{-1}$ is seen in **Figure 2C** as illustrated by the K48I variant at pH 8.5, but is also seen for WT and E47A (data not shown). This clearly indicates that an oxygen atom from the solvent is coordinated to the ferric site. When a $^2H_2O$ buffer is used, the 471 cm$^{-1}$ band downshifts by 6 cm$^{-1}$ to 465 cm$^{-1}$ indicating that the ligand possesses an easily exchangeable proton (**Figure 2C,c**); no other RR bands are seen to shift. The 471 cm$^{-1}$ frequency and isotopic shifts of –15 cm$^{-1}$ ($^{16}O/^{18}O$) and -6 cm$^{-1}$ ($^2H/H$) are completely compatible with a high-spin $Fe^{3+}$-OH species, as observed for cytochrome c oxidase (477 cm$^{-1}$),[29] methemoglobin and metmyoglobin (490-497 cm$^{-1}$)[30-32] and FixL (479 cm$^{-1}$).[33] Therefore we assign the 471 cm$^{-1}$ band to the high spin $Fe^{3+}$-OH stretching mode for which the frequency varies slightly for WT (466 cm$^{-1}$) and the E47A variant (467 cm$^{-1}$). The smaller $^{18}O$ isotopic shifts observed here for SOR compared to the other heme proteins[28-30] (-20/-25 cm$^{-1}$) indicates deviation from a pure $Fe^{3+}$-OH stretching mode. This could arise from coupling with other Fe-S modes and/or that the HO$^-$ species in SOR is involved in H-bonds at the solvent-exposed active site as is evidenced by the large bandwidth of the 466-471 cm$^{-1}$ band (ca. 40 cm$^{-1}$ FWHM).

The identification here of a high-spin $Fe^{3+}$-OH species is also totally consistent with the reported pH-induced spectral and redox changes of SOR active site.[19] In fact, the observed lowering of the redox midpoint potential of the SOR active site at basic pH [19] is consistent with the addition of the negatively charged HO$^-$ ligand. The spectral shift of the 644 nm CT band to 560 nm at high pH is consistent with contributions from an HO$^-$ → high spin $Fe^{3+}$ CT band; such CT bands at ~600 nm have been revealed by RR excitation profiles for high spin $Fe^{3+}$-OH species in myoglobin,[30-32] which enhances the Fe-OH vibrations as is observed here using 647.1 nm excitation. The resulting 560 nm band must be a mixture of electronic contributions of both the HO$^-$ → and CysS → high spin $Fe^{3+}$ CT bands since both the SOR Fe-OH and Fe-S modes are resonantly enhanced at 647.1 nm.

Another important observation in this work involves the differences observed for the Fe-OH stretching frequencies between WT and E47A on the one hand (466-467 cm$^{-1}$) and the K48I variant on the other (471 cm$^{-1}$). The global RR spectra of these SOR proteins are identical, which indicates that the vibrational structure of the Fe-S moiety of the active site was not perturbed by the K48I mutation. Thus, it seems unlikely that the change in Fe-OH bond strength is arising from changes in the Fe-S bond strength (i.e. back-donation). The change in the Fe-OH vibrational frequency of the K48I variant must arise from differences in H-bonding or electrostatic interactions with the ligand. For the case of the WT and E47A variant, the positively charged lysine side chain may be in H-bonding interaction (directly or indirectly via a H-bonding network through the solvent) with the HO$^-$ ligand. Such a specific interaction would not exist in the K48I variant. It is not clear yet from available crystal structures whether lysine 48 can make a direct H-bond to an HO$^-$ ligand, nevertheless, due to the fact that the HO$^-$ ligand is solvent-exposed, at least one water molecule could bridge the lysine 48 and the HO$^-$ ligand. The resulting polarized H-bond donated by the positively charged lysine residue will draw electron density from the Fe-OH bond and weaken it. Upon mutation to an electrically neutral isoleucine side chain, this polarized H-bond is lost, resulting in a strengthened Fe-OH bond, i.e. the 5 cm$^{-1}$ increase in its observed frequency for the K48I variant.

With respect to the E47A variant, the Fe-OH stretching frequency is almost identical to that of the WT. As underlined above, a similar polarized H-bond network involving the HO$^-$ ligand and the K48 residue would be also present for the case of the E47A variant. In the WT, below pH 9.0, the sixth coordination site is occupied by the E47 residue.[23] As shown here, pH values above 9.0 are required for the HO$^-$ ligand to displace E47 as sixth ligand in WT. Then, mutation of the E47 residue is expected to favor binding of the OH$^-$ group at more acidic pH, hence decreasing the pK$_a$ of the transition.

A Fe$^{3+}$-OH species has been implicated in the activity of superoxide dismutase but has never been unambiguously spectroscopically identified[34] and for the first time here is observed for SOR. Curiously, although a similar alkaline optical transition has been observed for the SOR from *P. furiosus*,[25] no RR evidence for a Fe$^{3+}$-OH species was observed. This might be due to differences in the relative resonance Raman enhancements of the Fe-S and Fe-OH modes in its active site.[25]

To investigate the reasons why a Fe$^{3+}$-OH species was not observed for *P. furiosus*, we have studied the pH dependence of the RR spectrum of the SOR from *Treponema pallidum*,[10,12,35] another SOR protein which, like *P. furiosus*[14], does not possess the additional rubredoxin-like Fe$^{3+}$(SCys)$_4$ center.

The RR measurements have been performed on the *T. pallidum* E48A variant (homologous to the *D. baarsii* E47A variant) at pH 5 and pH 9 between which the alkaline transition for this SOR occurs[35]. The WT SOR from *T. pallidum* also exhibits an alkaline optical transition from 650 to 560 nm above pH 8.5 and like the *D. baarsii* E47A SOR,[19,35] the E48A mutation in *T. pallidum* results in a decrease of the $pK_a$ of the optical alkaline transition to a value of 6.0.[35] The pH dependent RR spectra of the *T. pallidum* E48A variant are shown in **Figure 6**. For *T. pallidum*, we observed a very weak 473 cm$^{-1}$ which was sensitive to $H_2^{18}O$ experiments, shifting by –23 cm$^{-1}$ to 450 cm$^{-1}$. Thus, for *T. pallidum*, a similar Fe-OH species is also formed and observed in the active site at high pH, but with very weak resonance enhancement when compared to that observed for the *D. baarsii* enzyme. A similar very weak resonance enhancement behavior may explain why a $Fe^{3+}$-OH species could not be identified for the *P. furiosus* SOR.[25]

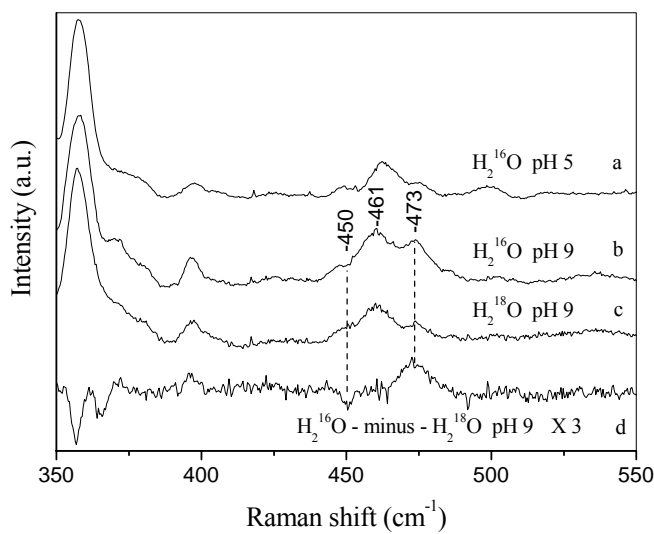

**Figure 6.** Low temperature (15 K) resonance Raman spectra of *Treponema pallidum* E48A SOR (3 mM) variant in the 350-550 cm$^{-1}$ range. Same Raman conditions as in Figure 1. Trace (a), protein in 150 mM acetate / $H_2^{16}O$ buffer pH 5.0. Trace (b) protein in 150 mM glycine NaOH / $H_2^{16}O$ buffer pH 9.0. Trace (c) same as spectrum b except in $H_2^{18}O$ buffer. Trace (d) Difference spectrum of [trace (b) $^{16}O$ – minus – $^{18}O$ trace (c)] X 3. Difference spectrum was calculated by cancellation of the 460 cm$^{-1}$ Raman band of the active site.

**Implications for SOR mechanism.**

In a previous paper, we proposed that the protonated form (BH) of the base (B⁻) responsible for the optical absorption and redox alkaline transitions could be involved in the second protonation process during the SOR catalytic cycle.[19] In that work, it was demonstrated that neither the conserved Glu47 and Lys48 residues of the *D. baarsii* SOR were directly responsible for the alkaline transition, however the identity of BH (whether another amino acid residue or a water molecule) could not be unambiguously determined. Here we clearly identified this base as an HO⁻ ligand on the ferric iron active site and consequently its protonated form as a water molecule. Furthermore, we have presented RR evidence that Lys48 is involved in a specific H-bonding interaction with the HO⁻ ligand and it is reasonable to expect that a similar Lys48 H-bonding interaction is exerted on the corresponding water molecule at the active site. . In the case of the *D. baarsii* enzyme, following the initial reaction of the ferrous iron site with superoxide, the first intermediate, a ferric iron-peroxo species, becomes protonated to form a ferric iron hydroperoxo species. This protonation process was shown to arise from the bulk solvent.[19] During this first protonation process, deprotonation of BH was not expected, because the corresponding formation of B⁻, i.e. $Fe^{3+}$-OH, can not occur until the intermediate species has left the ferric iron site. Involvement of BH ($H_2O$) in the second protonation process involving the ferric-hydroperoxo intermediate species would then allow the release of $H_2O_2$ together with the formation of the $Fe^{3+}$-OH species (**Figure 7**).

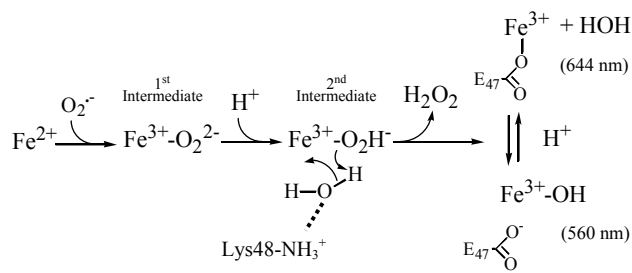

**Figure 7.** A proposed mechanism for the *D. baarsii* SOR based on Ref. 19 and this work. The values in parentheses indicate the wavelength of the electronic absorption maxima of the active site high spin $Fe^{3+}$ center at the end of the SOR reaction. The 644 nm species corresponds to the SOR in the final oxidized $Fe^{3+}$ high spin state with glutamate 47 at the sixth coordination position while the 560 nm species corresponds to the active site with HO⁻ ligand; the "alkaline transition" corresponds to conversion from the former to the latter. During SOR reaction, protonation of the second intermediate ($Fe^{3+}$-OOH) could most likely come from a water molecule for which the acidity is increased by H-bonding interactions associated with Lys48 (see text). The resulting HO⁻ anion will then compete with $H_2O$, buffer and Glu 47 for iron coordination after $H_2O_2$ is released.

In fact, the acidity of the water molecule would be increased because of the direct or indirect H-bond interaction associated with lysine 48. After release of $H_2O_2$, the $Fe^{3+}$-OH would be readily protonated, according to the pH of the surrounding bulk water solvent. It should be noted that the rate constant of the last step in the SOR reaction, namely the release of $H_2O_2$, could not be directly measured in the case of the *D. baarsii* enzyme, but was estimated to be slow, less than 5 $s^{-1}$.[19] These points clearly require further studies, in particular to determine the dependence of this second protonation process versus the pH.

**AKNOWLEDGEMENTS**

This work was in part supported by grant (TAM) from the Regional Council of the l'Ile-de-France (S.E.S.A.M.E.).

# Table of Contents Graphic

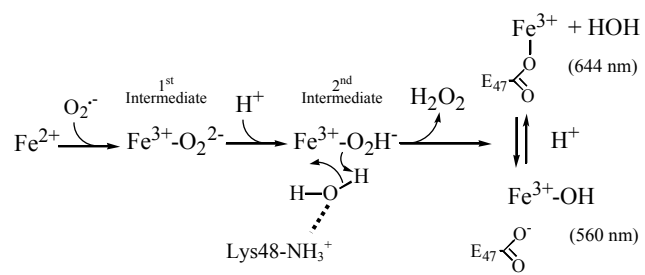